\documentclass[aps,
prd, 
twocolumn,
nofootinbib,superscriptaddress,floatfix
]{revtex4-1}

\usepackage{natbib}

\usepackage{graphicx}
\usepackage{dcolumn}
\usepackage{subfigure}
\usepackage{epstopdf}

\usepackage{amsfonts}
\usepackage{amsmath}
\usepackage{ifpdf}
\usepackage{comment}
\pdfoutput=1 

\newif\ifpdf
\ifx\pdfoutput\undefined
\pdffalse 
\else
\pdfoutput=1 
\pdftrue
\fi

\def\kt{$k_T$}
\newcommand{\nb}{{\bar n}}

\renewcommand{\log}{\ln}
\def\kp{k_3^+}
\def\km{k_3^-}
\newcommand{\ptm}{p_3^-}

\def\mo{{\mu_0}}

\def\dsdy{\frac{1}{\sigma_0}\frac{d\sigma_\text{\kt}^\text{dijet}}{dy_c}}
\def\dsdyinline{({1}/{\sigma_0})({d\sigma_\text{\kt}^\text{dijet}/}{dy_c})}
\def\S{{\bf\mathcal{S}}}
\def\g{\tilde{\gamma}}
\def\one{{(1)}}

\newcommand{\al}{\alpha}

\newcommand{\ep}{\epsilon}

\newcommand{\si}{\sigma}

\newcommand{\om}{\omega}

\def\be{\begin{equation}}
\def\ee{\end{equation}}
\def\ba{\begin{eqnarray}}
\def\ea{\end{eqnarray}}
\def\bfig{\begin{figure}[t]}
\def\efig{\end{figure}}

\newcommand{\nn}{\nonumber \\}
\def\eqn#1{(\ref{#1})}
\def\fig#1{Fig.\ \ref{fig:#1}}
\def\sec#1{Sec.\ \ref{sec:#1}}
\def\tab#1{Table\ \ref{tab:#1}}
\def\bwtxt{\begin{widetext}}
\def\ewtxt{\end{widetext}}

\makeatletter
\def\fmslash{\@ifnextchar[{\fmsl@sh}{\fmsl@sh[0mu]}}
\def\fmsl@sh[#1]#2{%
   \mathchoice
     {\@fmsl@sh\displaystyle{#1}{#2}}%
     {\@fmsl@sh\textstyle{#1}{#2}}%
     {\@fmsl@sh\scriptstyle{#1}{#2}}%
     {\@fmsl@sh\scriptscriptstyle{#1}{#2}}}
\def\@fmsl@sh#1#2#3{\m@th\ooalign{$\hfil#1\mkern#2/\hfil$\crcr$#1#3$}}

\makeatother

\begin{document}
\ifpdf
\DeclareGraphicsExtensions{.pdf, .jpg}
\else
\DeclareGraphicsExtensions{.eps, .jpg}
\fi




\title{The Exclusive \kt\ Dijet Rate in SCET with a Rapidity Regulator}
\author{William Man-Yin Cheung}
\email{mycheung@physics.utoronto.ca}
\affiliation{Department of Physics, University of Toronto, 
     60 St.\ George Street, Toronto, Ontario, Canada M5S 1A7}
\author{Simon M. Freedman}
\email{sfreedma@physics.utoronto.ca}
\affiliation{Department of Physics, University of Toronto, 
     60 St.\ George Street, Toronto, Ontario, Canada M5S 1A7}
\date{\today}

\begin{abstract}
We study the (exclusive) \kt\ jet algorithm using effective field theory techniques.  Regularizing the virtualities and rapidities of graphs in the soft-collinear effective theory (SCET), we are able to write the next-to-leading-order dijet cross section as the product of separate hard, jet, and soft contributions.  We show how to reproduce the Sudakov form factor to  next-to-leading logarithmic accuracy previously calculated by the coherent branching formalism.  Our result only depends on the renormalization group evolution of the hard function, rather than on that of the hard and jet functions as is usual in SCET.  We comment that regularizing rapidities is not necessary in this case.
\end{abstract}

\maketitle


\section{Introduction}\label{sec:intro}

Jets are important for understanding the background to new physics being investigated at the Large Hadron Collider.  Jet production is a multi-scale process that involves the large energy of the jet, $Q$, and its small invariant mass, $m_\text{jet}$.  A hierarchy of scales $Q\gg m_\text{jet}$ gives rise to large logarithms of the form $L\equiv\log(Q^2/m_\text{jet}^2)\gg1$ in perturbative calculations.  These logarithms manifest in the jet production rate in the form
\be\label{PTexp}
	R=\sum_{n=0}^\infty\sum_{m=0}^{2n}R_{nm}\al_s^n L^m,
\ee 
where $\al_s$ is the strong coupling constant.  Even when $\al_s\ll1$, the large logarithms will ruin perturbation theory when $\al_sL^2\sim1$. 

Well-known perturbative QCD (pQCD) techniques based on factorization theorems \cite{Collins:1989gx} and the coherent branching formalism \cite{Catani:1992ua} can sum these logarithms by writing the series \eqn{PTexp} as
\ba\label{csigma}
	R= C(\al_s)\Sigma(\al_s,L),
\ea
where
\ba\label{sudakov} 
	C(\al_s)&=& \sum_{n=0}^\infty C_n\al_s^n,\\
	\log\Sigma(\al_s,L)=L f_0(\al_s L)+&f_1&(\al_s L)+\al_s f_2(\al_s L)+\ldots\,.\nonumber
\ea 
The coefficient function $C(\al_s)$ contains no large logarithms $L$, while $\Sigma(\al_s,L)$ sums the logarithms.  The $f_0$ term sums the leading logarithms (LL), the $f_1$ term sums the next-to-leading logarithms (NLL), and the $f_{n\geq2}$ terms sum the subleading logarithms.  In this paper we will always refer to the logarithmic order in the exponent \eqn{sudakov} as opposed to the logarithmic order in the perturbative rate \eqn{PTexp}.

An example of a jet definition is the (exclusive) \kt\ jet algorithm \cite{Catani:1991gn,Catani:1991hj}, proposed to resolve the exponentiation issue of the earlier JADE algorithm \cite{Catani:1991gn,Brown:1990nm,Brown:1991hx}.  The \kt\ and JADE algorithms combine final-state partons into jets using a distance measure $y_{ij}$ for all pairs of final-state partons $\{i,j\}$.  If the smallest $y_{ij}$ is smaller than some pre-determined resolution parameter $y_c$, then that pair of partons are combined and all the $y_{ij}$'s are re-calculated.   The procedure is repeated until all $y_{ij}>y_c$, and these pseudo-partons are then called jets. The \kt\ algorithm measure for  $e^+e^-\to\ $ jets is
\ba\label{massiveyij}
	y_{ij}=2(1-\cos\theta_{ij})\frac{\min(E_i^2,E_j^2)}{Q^2}
\ea
where $Q$ is the centre-of-mass energy, $\theta_{ij}$ the angle between the final-state pair, and $E_{i,j}$  their respective energy.   We are interested in a two-jet final state where the cut parameter is small.  Jets in the $y_c\ll1$ region have small mass $m_\text{jet}\approx\sqrt{y_c}Q\ll Q$, which gives rise to large logarithms $L\equiv\log(1/y_c)$.  The \kt\ dijet production rate has been calculated using the coherent branching formalism to full LL accuracy in \cite{Catani:1991gn,Catani:1991hj} and partial NLL accuracy in \cite{Dissertori:1995qx}.  Clustering effects among multiple gluon emissions generate unsummed logarithms that start at $O(\alpha_s^2L^2)$ in the exponent \cite{Banfi:2001bz} and ruin the NLL summation of \cite{Dissertori:1995qx}.  

Effective field theory (EFT) techniques offer another approach to summing the large logarithms.   Using EFTs has the advantage of using the renormalization group (RG) to sum the large logarithms, as well as providing a systematic approach to power corrections.   In \cite{Cheung:2009sg} the \kt\ dijet rate was calculated using soft-collinear effective theory (SCET) to next-to-leading order (NLO).  SCET \cite{Bauer:2000ew,Bauer:2000yr,Bauer:2001ct, Bauer:2001yt,Bauer:2002nz,Freedman:2011kj} describes QCD using highly boosted ``collinear'' fields  and low energy ``soft'' fields.  SCET has previously been  successful in calculating jet shapes \cite{Hornig:2009vb, Ellis:2010rwa}, where it automatically separated the hard scattering interaction from the highly boosted interactions in the jets and from the soft radiation between them.    Such a separation allows the rate to be written as the convolution of hard, jet (one for each of the dijets), and soft functions,
\ba\label{hjs}
R=H\times J\times\bar J\times S,
\ea
each of which depends on a different scale.  These functions are then run individually to a common scale for logarithm summation.  The authors of \cite{Cheung:2009sg}, however, were unable to use dimensional regularization to regulate the individual NLO collinear and soft graphs of the \kt\ dijet rate, making it unclear how to write the rate as separate jet and soft functions as in \eqn{hjs}.

Recently \cite{Chiu:2011qc,Chiu:2012ir} a new regulator capable of regulating these divergences has been proposed.  This new ``rapidity regulator'' effectively places a cut on the rapidities of the fields \cite{Chiu:2012ir}, enabling the rate to be written as separate scheme dependent jet and soft functions.     The  rapidity regulator was used to sum logarithms in the jet broadening event shape \cite{Hornig:2009vb,Chiu:2011qc,Chiu:2012ir}, which has a similar issue at NLO to the \kt\ dijet rate.  The introduction of the rapidity regulator opens up the possibility of the RG running in another scale $\nu$, in analogy to the usual RG running scale $\mu$ of dimensional regularization. 

We propose to extend the work of \cite{Cheung:2009sg} using the new rapidity regulator and investigate how to write the \kt\ dijet rate as the product of hard, jet, and soft functions as in \eqn{hjs}.  Our work provides another application of the rapidity regulator.    As in \cite{Catani:1991gn,Catani:1991hj,Dissertori:1995qx} we assume a factorization theorem, which allows us to interpret the SCET collinear and soft graphs as the jet and soft functions that are run using the RG.  We can then use the RG to attempt to sum the large logarithms.  We find that we reproduce the coherent branching formalism result \cite{Dissertori:1995qx} but that neither approach sums the logarithms generated by clustering effects \cite{Banfi:2001bz}.  A similar result was recently found for the inclusive \kt\ algorithm \cite{Kelley:2012kj}.

The summation of the logarithms in the \kt\ dijet rate using SCET only requires the running of the hard function to NLL accuracy.  The jet and soft functions act as a single soft function $\S = J\times\bar J\times S$ that reproduces the infrared physics of QCD and depends only on a single soft scale. For NLL accuracy, it is unnecessary to define separate scheme-dependent jet and soft functions using the rapidity regulator.  

The rest of the paper proceeds as follows: in \sec{previous} we review the NLO results and issues of \cite{Cheung:2009sg}, and in \sec{lo} we show how the rapidity regulator solves these issues.  In \sec{nll} we show our final result with NLL summation and compare with the coherent branching formalism result.   We discuss the interpretation of our results and the utility of the rapidity regulator in \sec{discuss}. We conclude in \sec{conc}.

\section{Review of Previous Work}\label{sec:previous}

The \kt\ algorithm was previously studied using SCET in \cite{Cheung:2009sg}.  SCET is the appropriate EFT to describe QCD with highly boosted massless fields.  Collinear fields describe the boosted particles, and soft fields describe the low-energy particle exchanges.   The interactions within each sector (soft, collinear in each direction) decouple from one another and are described by a copy of QCD \cite{Freedman:2011kj}.  The interactions between sectors in the full theory are reproduced in the currents via Wilson lines \cite{Bauer:2000yr,Bauer:2001ct, Bauer:2001yt,Bauer:2002nz,Freedman:2011kj}.  

The appropriate SCET operator for dijet production where $n$ and $\nb$ are respectively the light-like directions of the jets  is \cite{Bauer:2002nz, Bauer:2006mk}
\ba\label{o2}
	O_2=\left[\bar\xi_nW_n\right]\left[Y_n^\dagger\Gamma Y_\nb\right]\left[W_\nb^\dagger\xi_\nb\right]
\ea
where $\xi_{n,\nb}$ is a two-component $n$- or $\nb$-collinear spinor.  The Wilson lines are defined in momentum space as
\ba\label{wilson}
	W_n &=& \sum_{\rm perm}\left[{\rm exp}\left( \frac{-g}{\bar{n}\cdot\mathcal P}\nb\cdot A_n \right )\right] \nn
	Y_n &=& \sum_{\rm perm}\left[{\rm exp}\left( \frac{-g}{n\cdot \mathcal P}n\cdot A_s \right )\right],
\ea
with $W_\nb$ and $Y_\nb$ defined analogously.  Here $\mathcal P^\mu$ is the momentum operator that acts on the gluon fields.  The fields $A_s, A_n,$ and $A_\nb$ represent soft, $n$-, and $\nb$-collinear gluon fields respectively.  The matching between QCD and SCET is well known \cite{Manohar:2003vb} and gives the matching coefficient
\ba\label{c2}
	C_2(\mu)&=&1+\frac{\al_s C_F}{2\pi}\left(-\frac12\log^2\frac{\mu^2}{-Q^2}-\frac32\log\frac{\mu^2}{-Q^2}-4+\frac{\pi^2}{12}\right) \nn
	&&+\ldots
\ea
and $\overline{\text{MS}}$ counterterm
\ba\label{z2}
	Z_2(\mu)=1+\frac{\al_s C_F}{2\pi}\left(\frac1{\ep^2}+\frac{3}{2\ep}+\frac1\ep\log\frac{\mu^2}{-Q^2}	\right)+\ldots.
\ea
The ellipses denote higher orders in $\al_s$.  The matching coefficient reproduces the UV physics of QCD.  

The $e^+e^-\to\gamma^\ast\to$ dijet rate is calculated in SCET by summing the collinear and soft diagrams and integrating over the appropriate phase space.  Generally the rate is written in the form
\ba\label{dsdy}
	\frac{1}{\sigma_0}\frac{d\sigma^\text{dijet}}{dy_c}=H\times(J\times\bar J\times S)
\ea
where $\si_0=(4\pi\al^2/Q^2)\sum_f e_f^2$ is the Born cross section\footnote{For dijet rates via a $Z^0$, only the Born cross section is modified. This is irrelevant for our calculation.}.  The soft contribution $S$ describes the interaction of the soft fields, while the hard function $H$ captures the physics of the hard initial interaction.  The hard function is defined to be $H=|C_2|^2$.  The jet contributions $J$ and $\bar J$ describe the interactions of the $n$- and $\nb$-collinear fields respectively.

\bfig
	\centering
	\subfigure[]{\includegraphics[width=0.3\columnwidth]{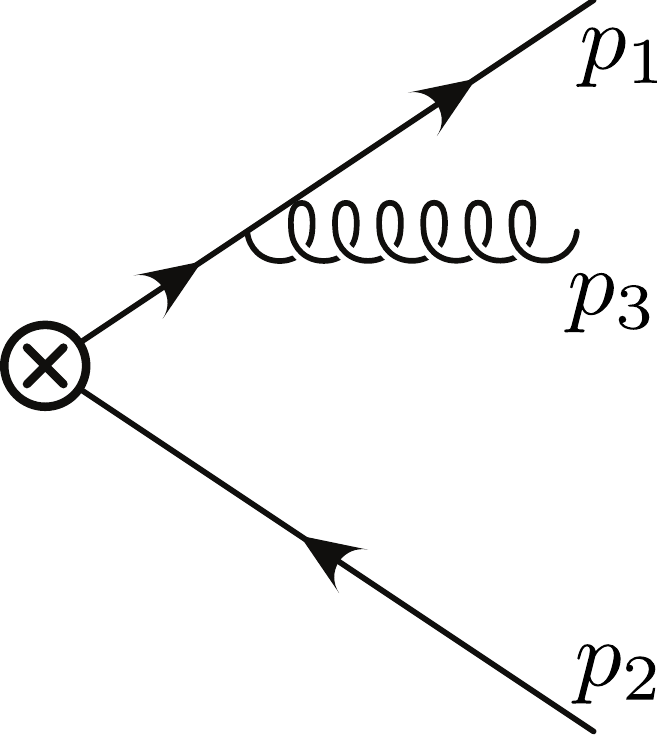}}\hspace{0.05\textheight}
	\subfigure[]{\includegraphics[width=0.3\columnwidth]{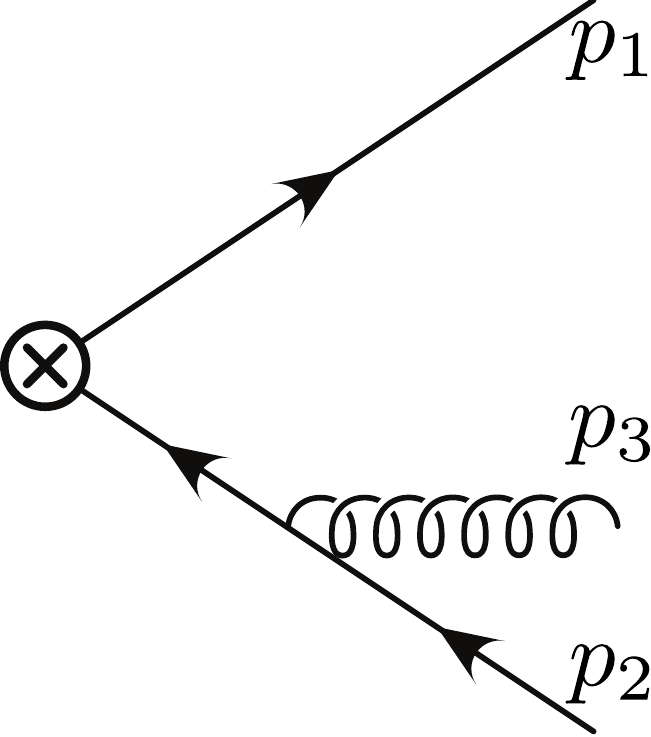}}
	\caption{QCD diagrams for real emission}
	\label{fig:dijet}
\efig

For perturbative calculations, the contributions in \eqn{dsdy} are individually written as
\ba
	F(\mu)=1+F^\one(\mu)+F^{(2)}(\mu)+\ldots
\ea
where $F=H,J, \bar{J}, S$ and $F^{(n)}$ is the $O(\al_s^n)$ term. The two QCD diagrams that contribute to real emission at NLO are shown in \fig{dijet}.  In SCET, the gluon can either be soft, $n$-, or $\nb$-collinear, resulting in six graphs that must be summed.  We write all momenta in lightcone coordinates $q^\mu=(n\cdot q, \nb\cdot q,\vec q_\perp)\equiv(q^+,q^-,\vec q_\perp)$.  We adopt the convention of \cite{Cheung:2009sg} and use the symbol $k\ll Q$ for soft momentum, and $p\sim Q$ for large momentum.  Contributions from the NLO collinear and soft graphs in dimensional regularization are given by integrating the corresponding differential cross sections over the relevant phase space $PS_F$ \cite{Cheung:2009sg}
\ba\label{functions}
	S^\one(\mu)&=& \frac{\al_s C_F}{2\pi} f_{\ep} \int_{PS_S}d\kp d\km \frac{2}{(\kp\km)^{1+\ep}}\\
	\tilde J^\one(\mu)&=&\frac{\al_s C_F}{2\pi}f_\ep\int_{PS_n}d\kp d\km \frac{(k_3^+p_3^-)^{-\epsilon}}{Qk_3^+}\\
	&&\qquad\times\left[ \frac{p_3^-}{Q}(1-\epsilon) + 2\frac{Q - p_3^-}{p^-_3}\right]\nonumber\\
	J^\one_{0}(\mu)&=&2\frac{\al_s C_F}{2\pi}f_\ep\int_{PS_{0}}\frac{d\kp d\km}{(\kp\km)^{1+\epsilon}}	
\ea
where $\tilde J$ is referred to as the naive collinear graph and $J_0$ the zero-bin. The ``true'' collinear contribution requires a zero-bin subtraction \cite{Manohar:2006nz} and is defined as $J(\mu)=\tilde J(\mu)-J_0(\mu)$. We have introduced $f_\ep\equiv \mu^{2\ep}e^{\ep\gamma_E}/\Gamma(1-\ep)$ for later convenience.  The $\nb$-collinear graph is the same as the $n$-collinear graph at NLO, $\bar J^\one(\mu)=J^\one(\mu)$.

\begin{figure*}[ht!]
	\centering
	\includegraphics[width=\textwidth]{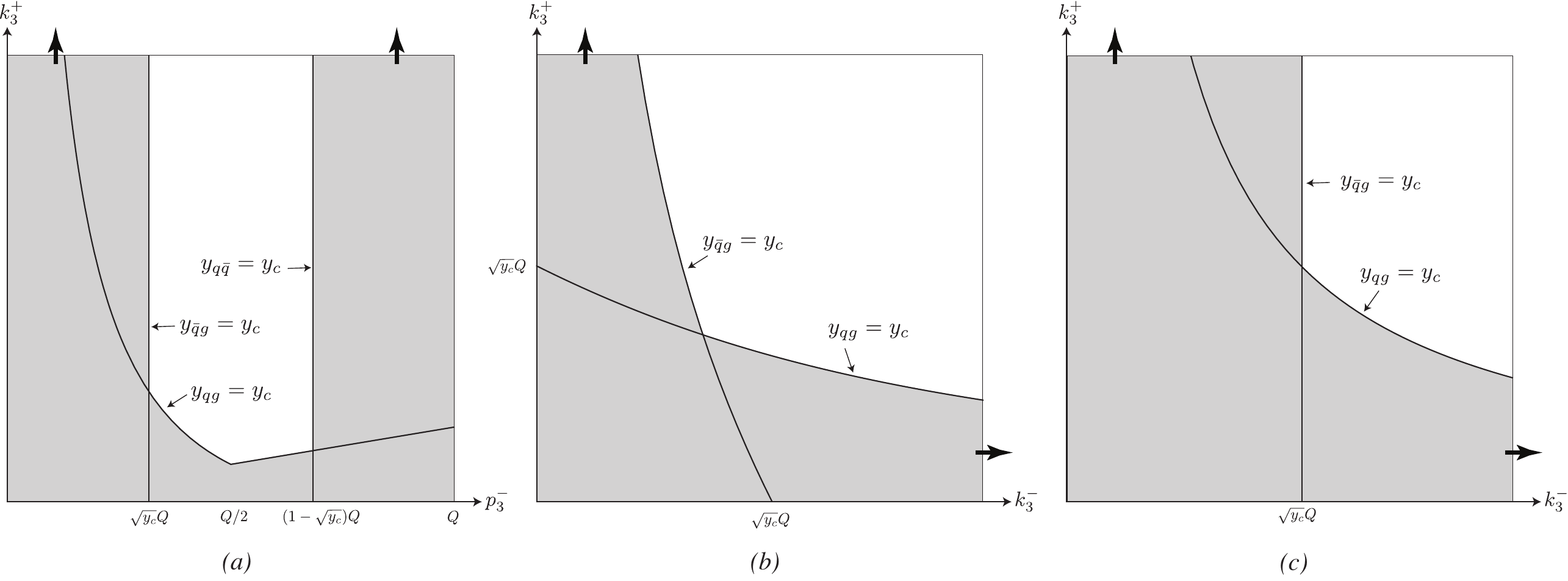}
	\caption{NLO \kt\ dijet phase space for (a) $n$-collinear gluon (b) soft gluon and (c) zero-bin.  These plots are taken from \cite{Cheung:2009sg}, and the bold arrows indicate that the plots extend to infinity. \label{fig:ktplot}}
\end{figure*}
\begin{table}[t]
\begin{tabular}{c|c|c}
	$n$-collinear & zero-bin & soft \\\hline
	$\min\left(\frac{\kp}{\ptm},\frac{\kp\ptm}{(Q-\ptm)^2}\right)<y_c$ & $\kp\km<y_c Q^2$	& $\kp(\kp+\km)<y_c Q^2$	\\
	 $\ptm<\sqrt{y_c}Q $ & $\km<\sqrt{y_c}Q$	&  $\km(\kp+\km)<y_c Q^2$\\
	 $\ptm>Q(1-\sqrt{y_c})$ & &
\end{tabular}
	\caption{Phase space constraints for NLO real emission for the \kt\ algorithm.  The constraints are plotted in \fig{ktplot}. \label{tab:constraints}}
\end{table}

The relevant NLO phase space constraints in SCET are found by applying the \kt\ measure \eqn{massiveyij} to the $q\bar{q}g$ final state and expanding in $k\ll p,Q$.  At leading order in power counting the fermions must be collinear, and we define $n^\mu$ to be in the direction of the quark.   The constraints for a soft and $n$-collinear gluon are shown in \tab{constraints} and plotted in \fig{ktplot}.  The constraints for an $\nb$-collinear gluon are the same as those for an $n$-collinear gluon with ``+'' and ``$-$'' interchanged. 

In \cite{Cheung:2009sg} it was found that the NLO soft graph can be written as 
\ba\label{softproblem}
	S^\one(\mu)&=&-2\frac{\al_s C_F}{\pi}\frac{e^{\ep\gamma_E}}{\ep\Gamma(1-\ep)}\left(\frac{\mu^2}{y_c Q^2}\right)^{\ep}\int_0^{1} dx\,\frac{(1-\frac{x^2}2)^{-\ep}}{x}\nn
	&&+\ldots,
\ea 
where the ellipses denote terms that are properly regulated in dimensional regularization.  The integral in \eqn{softproblem} is not regularized as $x\to0$, and this means that interpreting the soft function as the sum of the soft graphs as in \eqn{dsdy} is not well defined.  However, it was noted in \cite{Cheung:2009sg} that the NLO zero-bin can be written as
\ba\label{zbproblem}
	J^\one_{0}(\mu)=-\frac{\al_s C_F}{\pi}\frac{e^{\ep\gamma_E}}{\ep\Gamma(1-\ep)}\left(\frac{\mu^2}{y_c Q^2}\right)^{\ep}\int_0^1 \frac{dx}{ x}+\ldots,\nn
\ea
where again the ellipses denote terms that are properly regularized.  The $x\to0$ divergence in this integral is the same as the soft graph.   Because $J^\one_0(\mu)$ enters into both $J(\mu)$ and $\bar J(\mu)$ with a relative minus sign compared to the soft graph, the total rate $\dsdyinline$ is properly regularized at NLO as expected.   

Decomposing the rate as separately regularized jet and soft functions as in \eqn{dsdy}, where $J$ and $S$ are respectively the collinear and soft graphs, is therefore not possible using pure dimensional regularization.  The issue of separately well-defined functions comes from how phase space is being divided between the collinear and soft graphs in this scheme.  The soft graph is being integrated over the region $k_3^{\pm}\to0$ while keeping $\kp\km\leq y_c Q^2$. This is a highly boosted region, and is more naturally associated with the jet function than the soft function.  The jet broadening rate has a similar issue in SCET as shown in \cite{Hornig:2009vb}.

As pointed out in \cite{Cheung:2009sg}, the soft graph can be regulated using a different scheme such as a cut-off regulator.  The cut-off regulator removes the contribution of the aforementioned region from the soft graph and regulates the integral in \eqn{softproblem}.  The jet broadening rate can also be regularized using a cut-off.  The cut-off regulator, however, is not very attractive as it is not gauge invariant, making it hard to run using the RG.  It is also unclear how to define it in the naive collinear calculation.

Another scheme also studied in \cite{Cheung:2009sg} is to use offshellness as an infrared regulator, while using dimensional regularization to regulate the UV.  Here, the small quark and anti-quark offshellness regulates the integrals in \eqn{softproblem} and \eqn{zbproblem}.  However, the resulting collinear and soft contributions -- including the virtual diagrams -- are not individually infrared finite, even though these infrared divergences cancel in the total NLO rate as expected.  Therefore it is again unclear how to interpret these as the jet and soft functions of \eqn{dsdy}.

In the next section we use the recently introduced rapidity regulator \cite{Chiu:2011qc,Chiu:2012ir} to separate the low energy theory into jet and soft functions associated with the collinear and soft fields respectively.

\section{Next-to-Leading-Order calculation}\label{sec:lo}

In this section we show how all the divergences in the phase space of the soft graphs are tamed with the introduction of the rapidity regulator \cite{Chiu:2011qc, Chiu:2012ir}. The rapidity regulator was used to solve the similar issue and sum the logarithms in jet broadening \cite{Chiu:2011qc, Chiu:2012ir}.  The regulator acts as an energy cut-off in a similar way that dimensional regularization acts as a cut-off on the mass scale of loop momenta \cite{Georgi:1994qn}.  The form is similar to dimensional regularization and also maintains gauge invariance \cite{Chiu:2012ir}, unlike a cut-off regulator.   We will show in this section that using the rapidity regulator splits the NLO collinear and soft graphs into separately finite pieces.  This allows us to interpret the jet and soft functions as the collinear and soft interactions respectively. 

The rapidity regulator modifies the momentum-space definition of the Wilson lines \eqn{wilson} to \cite{Chiu:2011qc,Chiu:2012ir}
\ba\label{rapidityW}
	W_n &=& \sum_{\rm perm}{\rm exp}\left( \frac{-g}{\bar{n}\cdot\mathcal P}\left[ w^2\frac{|\bar{n}\cdot \mathcal P|^{-\eta}}{\nu^{-\eta}}\bar{n}\cdot A_n\right] \right ) \nn
	Y_n &=& \sum_{\rm perm}{\rm exp}\left( \frac{-g}{n\cdot \mathcal P}\left[ w\frac{ |2\mathcal P^3|^{-\eta/2}}{\nu^{-\eta/2}}n\cdot A_s\right] \right ).
\ea
Here $\mathcal P^3$ pulls down the component of momentum in the spatial direction of the jet.  The new parameter $\eta$ acts similarly to $\ep$ in dimensional regularization.  The parameter $w$ counts the number of emissions from a Wilson line, and is taken to one as $\eta\to0$.  Implementing the rapidity regulator modifies the NLO collinear  and soft graphs to
\ba\label{rapsigma}
	S^\one(\mu,\nu)&=& 2 \frac{\al_s C_F}{2\pi} w^2f_\ep\int_{PS_s}\frac{d\kp d\km}{(k_3^+k_3^-)^{1+\ep}}\left| \frac{k_3^+ - k_3^-}{\nu}\right| ^{-\eta} \nn
	\tilde J^\one(\mu,\nu)&=& \frac{\al_s C_F}{2\pi}f_\ep\int_{PS_n}d\kp d\ptm\frac{(k_3^+p_3^-)^{-\epsilon}}{Qk_3^+}\left[ \frac{p_3^-}{Q}(1-\epsilon) \right.\nn
	&&\left.\quad+ 2w^2\frac{Q - p_3^-}{p^-_3}\left( \frac{p_3^-}{\nu}\right) ^{-\nu}\right] \\
	J^\one_0(\mu,\nu)&=& 2\frac{\al_s C_F}{2\pi}w^2f_\ep\int_{PS_{0}}\frac{d\kp d\km}{(k_3^+k_3^-)^{1+\epsilon}}\left( \frac{k_3^-}{\nu}\right) ^{-\eta}\nonumber.
\ea
Note that the phase space constraints $PS_F$ are not affected. The pure dimensional regularized functions are recovered in the $\eta\to0$ limit.  

Calculating the collinear and soft graphs is now straightforward.  As has been previously demonstrated \cite{Chiu:2012ir}, we must expand in $\eta$ before $\ep$.  As we are considering the $y_c\ll 1$ region, all terms subleading in $y_c$ are also suppressed.

The naive NLO collinear graph is
\ba
	\tilde J^\one(\mu,\nu)&=&\frac{\al_s C_F}{2\pi}\left(4w^2\left(1-\frac{\pi ^2}{12}-\log2\right) -\frac12 +\ln 2 \right.\nn
	&&\left.+\left(\frac1\ep-\log\frac{Q^2 y_c}{\mu ^2}\right)\left(w^2(2+\log y_c)-\frac12\right)\right).\nn
\ea
We leave in $w$ for now and will set it to one at the end.  The logarithms cannot be minimized at any one scale because we have not yet included the zero-bin subtraction.  The NLO zero-bin contribution is
\ba
	J^\one_0(\mu,\nu)&=&\frac{\al_s C_F}{2\pi}w^2\left( - \frac{2}{\epsilon\eta} + \frac{1}{\epsilon}\ln\frac{y_cQ^2}{\nu^2} + \frac{2}{\eta}\ln\frac{y_cQ^2}{\mu^2}\right.\nn
	&&\left. - \ln\frac{y_cQ^2}{\mu^2}\ln\frac{y_cQ^2}{\nu^2}\right).
\ea
Subtracting the zero-bin from the naive collinear graph gives the true (bare) collinear contribution
\ba\label{J}
	&&J^{B\one}(\mu,\nu)=\frac{\al_s C_F}{2\pi}\left(4w^2\left(1-\frac{\pi ^2}{12}-\log2\right) -\frac12 +\ln 2\right.\nn
	&&+\left.\left(\frac1\ep-\log \left(\frac{Q^2 y_c}{\mu ^2}\right)\right)\left(2 w^2\left(\frac{1}{\eta }+1-\frac12\log\frac{Q^2}{\nu ^2}\right) -\frac{1}{2}\right)	\right).\nn
\ea
The collinear logarithms can be minimized at $\mu_J=\sqrt{y_c}Q$ and $\nu_J=Q$.  The $\nb$-collinear contribution $\bar J(\mu,\nu)$ is exactly the same as $J(\mu,\nu)$ at this order in $\al_s$. 

The NLO soft graph can be calculated similarly. The extra $\eta$-dependent piece in \eqn{rapsigma} regulates the divergence of \eqn{softproblem}.  The NLO (bare) soft graph is
\ba\label{S}
	S^{B\one}(\mu,\nu)&&=\frac{\al_s C_F}{2\pi} w^2\left(\log^2\frac{y_cQ^2}{\mu^2}-\frac{\pi^2}{3}	\right.\\
	&&\left.+2\left(\frac1\ep-\frac2\eta+\log\frac{y_cQ^2}{\nu^2}\right)\left(\frac1\ep-\log\frac{y_cQ^2}{\mu^2}\right)	\right),\nonumber
\ea
where the logarithms are minimized at the scales $\mu_S=\sqrt{y_c}Q=\nu_S$.  Note that the dimensional regularization scale of the soft graph is equal to that of the collinear graph, $\mu_J=\mu_S$.

Putting the collinear and soft graphs together, as well as the matching coefficient \eqn{c2} and the counterterm \eqn{z2}, the \kt\ dijet rate is
\ba
	&&\dsdy=H(\mu)J(\mu,\nu)\bar J(\mu,\nu) S(\mu,\nu)\\
	&&=1+\frac{\al_s C_F}{2\pi}\left(-\log^2y_c-3 \log y_c + \frac{\pi ^2}{6} - 1 -6 \log2\right)+\ldots,\nonumber
\ea
which exactly reproduces the pQCD result \cite{Catani:1991gn, Dissertori:1995qx, Brown:1991hx}.  All the graphs must be evaluated at the same $(\mu,\nu)$.  Notice that the $\nu$ dependence must cancel between the collinear and soft graphs because $H$ is $\nu$-independent.  This is a general result and means that, when added together, the $\eta$ dependence of the $J, \bar{J}$ and $S$ counterterms must vanish \cite{Chiu:2011qc}.

We find that unlike in \cite{Cheung:2009sg}, we can define the jet and soft functions in \eqn{dsdy} as the collinear and soft interactions respectively.  In the next section we show how to sum the logarithms using the RG by running each function individually.  We then compare the summed expression to the coherent branching formalism result.

\section{Next-to-leading logarithm summation}\label{sec:nll}

We wish to calculate both $f_0$ and $f_1$ of \eqn{sudakov} to sum the logarithms and compare to \cite{Dissertori:1995qx}.  Because we have two UV regulators, the jet and soft functions now run through a two-dimensional $(\mu, \nu)$ space. 

The renormalized function $F$ is defined in terms of the bare function $F^{B}$ and counterterm $Z_F$ as $F^{B}=Z_F F$.   Therefore, the anomalous dimensions in the two directions of the $(\mu,\nu)$ space are found using
\ba\label{anomdim}
	\gamma_F^\mu(\mu,\nu)&=&-\left(\frac{\partial}{\partial\log\mu}+\beta(\al_s)\frac{\partial}{\partial\al_s}\right)\log Z_F\nn
	\gamma_F^\nu(\mu,\nu)&=&-\left(\frac{\partial}{\partial\log\nu}+\beta(w)\frac{\partial}{\partial w}\right)\log Z_F
\ea
where $F=H,J,S$.  The running of the coupling constant $\beta(\al_s)=-2\al_s\ep+O(\al_s^2)$ is well-known and $\beta(w)=-\eta w/2$ exactly \cite{Chiu:2012ir}.  The counterterms of the jet and soft functions are found from \eqn{J} and \eqn{S} to be
\ba
	Z_S&=&1+\frac{\al_sC_F}{2\pi}\left(\frac2{\ep^2}-\frac{4}{\ep\eta}+\frac{2}{\ep}\log\frac{\mu^2}{\nu^2}+\frac{4}{\eta}\log\frac{\mo^2}{\mu^2}\right)+\ldots\nn
	Z_J&=&1+\frac{\al_sC_F}{2\pi}\left(\frac2{\ep\eta}+\frac{3}{2\ep}-\frac1\ep\log\frac{Q^2}{\nu^2}-\frac2\eta\log\frac{\mo^2}{\mu^2}\right)+\ldots\nn
\ea
where we have set $w=1$, $\mo^2=y_cQ^2$, and the ellipses here denote higher orders in $\al_s$.   The hard function counterterm $Z_H^{-1}\equiv|Z_2|^2=Z_S Z_J^2$ as expected.  The NLO anomalous dimensions are
\ba \label{gamma}
	\gamma_S^\mu&=&\frac{2\al_s C_F}{\pi}\log\frac{\mu^2}{\nu^2}	\qquad \gamma_J^\mu=\frac{\al_s C_F}{\pi}\left(\frac32+\log\frac{\nu^2}{Q^2}\right)\nn
	\gamma_S^\nu&=&\frac{2\al_sC_F}{\pi}\log\frac{\mo^2}{\mu^2}	\qquad \gamma_J^\nu=-\frac{\al_sC_F}{\pi}\log\frac{\mo^2}{\mu^2}\nn
	&&\gamma_H^\mu\equiv\gamma_H=-\frac{\al_sC_F}{\pi}\left(3+2\log \frac{\mu^2}{Q^2}\right).
\ea
The hard anomalous dimension in the $\nu$ direction vanishes identically because $Z_2$ is $\nu$-independent.  For consistency in the running, we must have
\ba\label{gammacondition}
	-\vec\gamma_H=2\vec\gamma_J+\vec\gamma_S,
\ea
where $\vec\gamma_F = (\gamma_F^\mu, \gamma_F^\nu) = - \vec\nabla\log Z_F$ with $\vec\nabla\equiv\left(\mu\frac{d}{d\mu},\nu\frac{d}{d\nu}\right)$. From \eqn{gamma}, we see that these conditions are satisfied at NLO.  

The anomalous dimensions allow the functions to be run to any scale.  However, unlike in the usual case of only using dimensional regularization to regulate the UV, the hard, jet, and soft functions are now scalar functions defined over a two-dimensional $(\mu,\nu)$ space. Path independence of running is equivalent to the curl of $\vec\gamma_F$ vanishing.  This vanishing curl gives the condition
\ba\label{curl}
	\mu\frac{d}{d\mu}\gamma_F^\nu(\mu,\nu)=\nu\frac{d}{d\nu}\gamma_F^\mu(\mu,\nu),
\ea
which, along with \eqn{gammacondition}, must be satisfied to all orders in $\al_s$.  We show in the Appendix that the soft $\nu$ anomalous dimension can be written as
\ba\label{gammanu}
	\gamma^\nu_S(\mu)=\gamma^\nu_S(\mo)+\int_{\mo}^{\mu} \frac{d\mu'}{\mu'}\left(\nu\frac{d}{d\nu}\gamma_S^\mu(\mu',\nu)\right),
\ea
where the general form of the soft $\mu$ anomalous dimension is taken to be
\ba\label{generalgamma}
	\gamma_S^\mu(\mu,\nu)=\Gamma_S[\al_s(\mu)]\log\frac{\mu^2}{\nu^2}+\gamma_S[\al_s(\mu)].
\ea
Here $\Gamma_S$ is called the cusp anomalous dimension. The $\gamma^\nu_S(\mo)$ contains no logarithms and all the logarithmic dependency of $\gamma_S^\nu(\mu)$ is determined by the $\mu$ anomalous dimension. A similar expression to \eqn{gammanu} appears in \cite{Chiu:2012ir}. The hard anomalous dimension has a similar form as \eqn{generalgamma} with $\nu=Q$ \cite{Hornig:2009vb}. The jet anomalous dimension is completely constrained by the hard and soft anomalous dimensions from \eqn{gammacondition}.

We can use the above to solve the RG equations and sum the logarithms.  Each function $F(\mu,\nu)$ must be evolved from the scale that minimizes its logarithms ($\mu_F,\nu_F$) to a common scale.  The solution to the RG equations gives the running of each function
\ba\label{rge}
	F(\mu_2,\nu_2)=F(\mu_1,\nu_1)e^{\int_{\mu_1}^{\mu_2}\frac{d\mu}{\mu}\gamma_F^\mu(\mu,\nu_2)} e^{\int_{\nu_1}^{\nu_2}\frac{d\nu}{\nu}\gamma_F^\nu(\mu_1,\nu)}\nn
\ea
where we have chosen to run in $\nu$ first but path independence is guaranteed with the use of \eqn{gammanu}.  The summed rate for the path in \fig{generalrun} is therefore
\ba\label{R}
	\dsdy=&&H(\mu_H)J(\mu_J,\nu_J) \bar J(\mu_J,\nu_J) S(\mu_S,\nu_S)\nonumber\\
		&&\times e^{K_H(\mu_H,\mu_J)}\left(\frac{\mu_J}{Q}\right)^{\omega_H(\mu_H,\mu_J)}
		\left(\frac{\nu_S}{\nu_J}\right)^{\omega_S(\mu_J,\mo)} \nn
\ea
where we have run to a general $(\mu,\nu)$, and used the consistency equations \eqn{gammacondition} and path independence \eqn{gammanu} to write everything in terms of the hard and soft running. Terms subleading to NLL accuracy have been suppressed.  Because of path independence, we can choose any other path and get the same NLL terms.  The summed rate is both $\mu$- and $\nu$-independent, as expected.  
\bfig
   \centering
   \includegraphics[width=0.75\columnwidth]{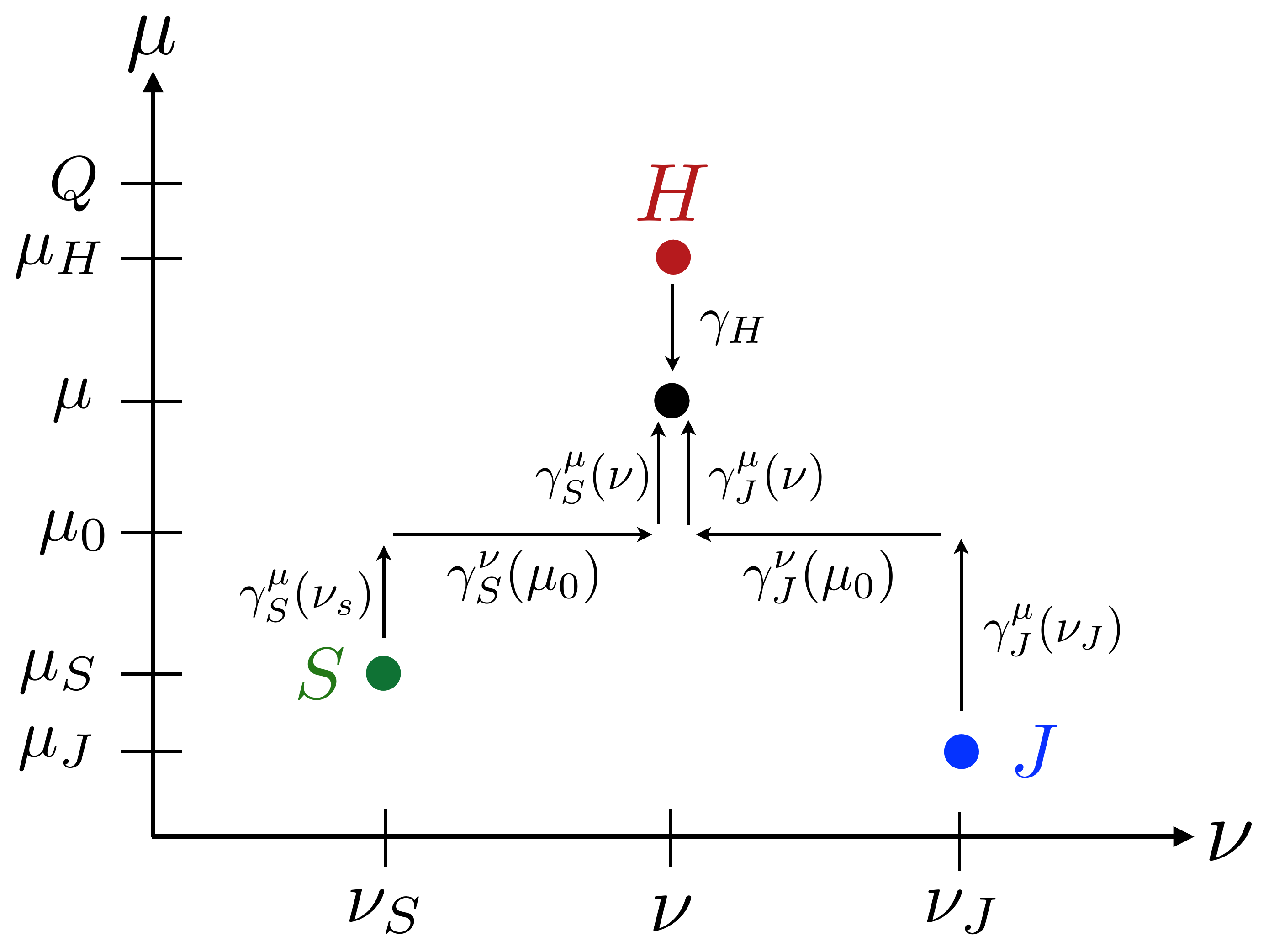} 
   \caption{Running each function from $(\mu_F,\nu_F)$ to $(\mu,\nu)$.}
   \label{fig:generalrun}
\efig

The running kernels in \eqn{R} are defined as
\ba\label{kernel}
	&\omega_F(\mu_1,\mu_2)&=-\frac{\Gamma_F^0}{\beta_0}\left[\log r+\left(K-\frac{\beta_1}{\beta_0}\right)\frac{\al_s(\mu_2)}{4\pi}(r-1)\right]\nn
	&K_F(\mu_1,\mu_2)&=-\frac{\gamma_F^0}{2\beta_0}\log r-\frac{2\pi\Gamma_F^0}{\beta_0^2}\left[\frac{r-1-r\log r}{\al_s(\mu_1)}\right. \\
	&&\left.+\left(K-\frac{\beta_1}{\beta_0}\right)\frac{1-r+\log r}{4\pi}+\frac{\beta_1}{8\pi\beta_0}\log^2r\right]\nonumber,
\ea
where we denote $r=\al_s(\mu_1)/\al_s(\mu_2)$.  The coefficients $\Gamma_F^n$ and $\gamma_F^n$ are given from the general form of the anomalous dimension \eqn{generalgamma} as
\ba\label{gammaexp}
	\Gamma_F[\al_s(\mu)]&=&\left(\frac{\al_s}{4\pi}\right)\Gamma_F^0+\left(\frac{\al_s}{4\pi}\right)^2\Gamma_F^1+\ldots\nn
	\gamma_F[\al_s(\mu)]&=&\left(\frac{\al_s}{4\pi}\right)\gamma_F^0+\left(\frac{\al_s}{4\pi}\right)^2\gamma_F^1+\ldots
\ea
where from \eqn{gamma} we can read off
\ba\label{G0g0}
	&\Gamma_H^0=-8C_F \qquad \quad &\gamma_H^0=-12 C_F\nn
	&\Gamma_S^0=8 C_F \qquad \quad  &\gamma_S^0=0.
\ea
The $\beta$-function of the coupling constant $\al_s$ also has an expansion
\ba
	\beta[\al_s(\mu)]=-2\al_s\left[\left(\frac{\al_s}{4\pi}\right)\beta_0+\left(\frac{\al_s}{4\pi}\right)^2\beta_1+\ldots\right]
\ea
where
\ba
	\beta_0&=&\frac{11C_A}{3}-\frac{2n_f}{3}\nn
	\beta_1&=&\frac{34 C_A^2}{3}-\frac{10C_An_f}{3}-2C_Fn_f.
\ea
The two-loop running in the coupling constant $\al_s(\mu)$ gives
\ba
	\frac{\al_s(Q)}{\al_s(\mu)}&=&1+\frac{\al_s(Q)\beta_0}{4\pi}\log\frac{\mu^2}{Q^2}\nn
	&&+\frac{\al_s(Q)\beta_1}{4\pi\beta_0}\log\left(1+\frac{\al_s(Q)\beta_0}{4\pi}\log\frac{\mu^2}{Q^2}\right).
\ea
The factor
\ba\label{cusp}
	K\equiv\left(\frac{67}{9}-\frac{\pi^2}{3}\right)C_A-\frac{10}{9}n_f
\ea
is the well-known ratio of the one- and two-loop cusp anomalous dimensions, $K=\Gamma_F^1/\Gamma_F^0$ \cite{Dissertori:1995qx, Hornig:2009vb}, and is required for the NLL summation.  

Choosing the scales that minimize the logarithms in the hard, jet, and soft functions
\ba\label{natural}
	\mu_H=Q& \qquad \quad &\mu_J=\mu_S=\mo\nn
	\nu_J=Q&  &\nu_S=\mo
\ea
simplifies \eqn{R} to
\ba\label{trueR}
	\dsdy&=&H(Q)J(\mo,Q)\bar J(\mo,Q) S(\mo,\mo)\nn
		&&\times e^{K_H(Q,\mo)}\left(\frac{\mo}Q\right)^{\omega_H(Q,\mo)},
\ea
which sums the rate to NLL accuracy.  From the above equation we can see that only the RG of the hard function is required for the summation to NLL accuracy.   The action of running in rapidity cancels between the jet and soft functions.  We will discuss this issue in more depth in the following section.

\begin{figure*}
	\centering
	\includegraphics[width=1.5\columnwidth]{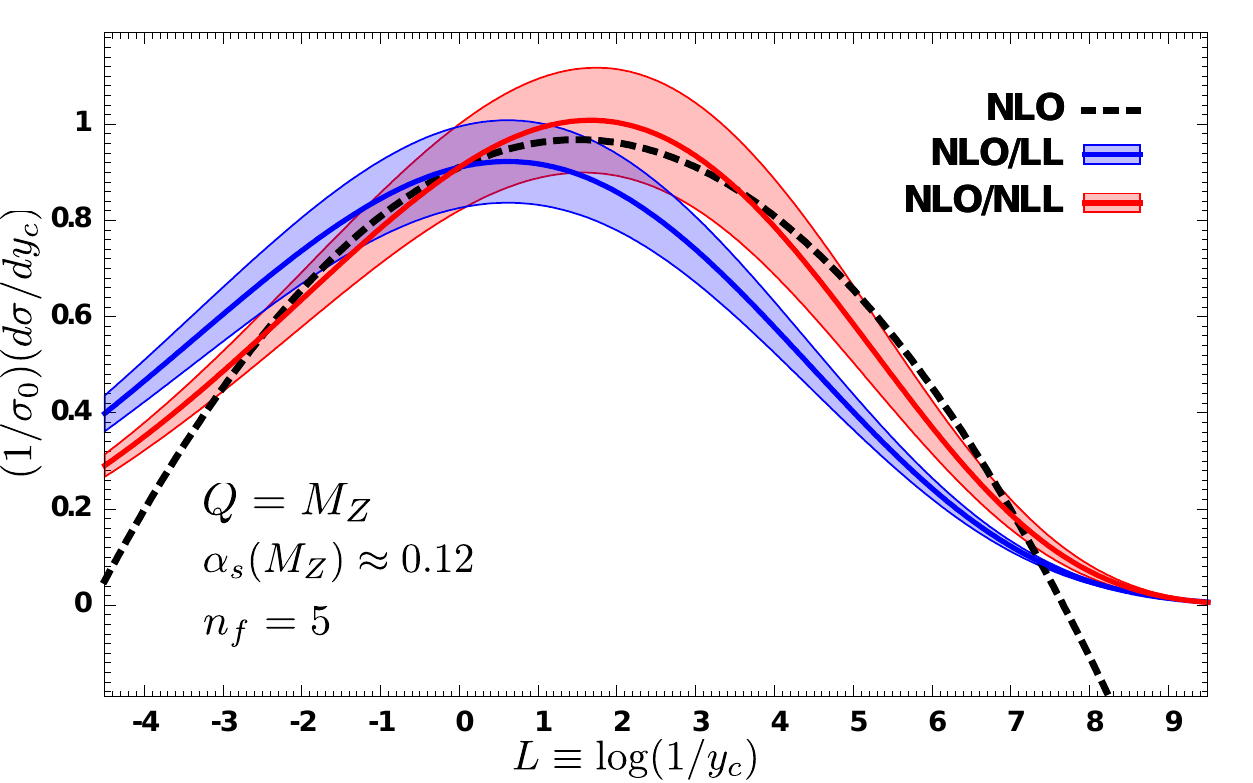}
	\caption{Plots of \eqn{trueR} for LL and NLL accuracy. NLO is the order in $\al_s$ the coefficient function $C(\al_s)$ of \eqn{sudakov} is taken to. The multipole expansion breaks down before $L\to0$ or $y_c\to1$, and the inclusion of NLO terms in $C(\al_s)$ improves accuracy of the curve over this region. The procedure for calculating the error bands is described in the text. }
	\label{fig:error}
\end{figure*}

We can now find the functions $f_0$ and $f_1$ of \eqn{sudakov} from \eqn{trueR}.  The LL summation comes from setting $\gamma_H^0=K=\beta_1=0$.  The NLL summation comes from the terms proportional to a single power of $\gamma_H^0$, $K$, and $\beta_1$. Therefore,
\ba
	f_0&=&-\frac{\Gamma_H^0}{2\beta_0}\left(1+\frac{\log(1-x)}x\right)	\\
	f_1&=&\frac{\gamma_H^0}{2\beta_0}\log(1-x)+\frac{\Gamma_H^0K}{2\beta_0^2}\left(\frac{x}{1-x}+\log(1-x)\right)\nn
	&&-\frac{\Gamma_H^0\beta_1}{2\beta_0^3}\left(\frac{x+\log(1-x)}{1-x}+\frac12\log^2(1-x)\right)\nonumber
\ea 
where $x\equiv \al_s C_F\beta_0 L/(4\pi)$.  Using \eqn{G0g0} we see that the functions agree exactly with the coherent branching formalism result \cite{Dissertori:1995qx}.   We plot the summed rate in \fig{error} as a function of $\log(1/y_c)$. The maximum jet production is at around $y_c\simeq0.2$, which corresponds to jets of mass $\sqrt{y_c} M_Z\simeq 40$GeV for LEP -- well above $\Lambda_\text{QCD}$.   

The error in \fig{error} is found by varying the scales $\mu_{H,J,S}$ and $\nu_{J,S}$ in \eqn{R} by $2$ and $1/2$ of their values in \eqn{natural}.  We vary the jet and soft scales together to maintain the $\mu_J\approx\mu_S$ scaling.  Varying the $\nu_F$ scales without varying $\mu_J$ produces no error due to the exponent $\om_S(\mu_J = \mo,\mo) = 0$.  We take a naive approach to estimate the correlated errors by varying $\mu_J$ and $\nu_F$ together, and taking the geometric mean of the resulting percent errors.

\section{Discussion}\label{sec:discuss}

That only the RG of the hard function is necessary for NLL accuracy suggests the \kt\ dijet rate should be written as
\ba\label{hs}
	\dsdy=H(\mu)\S(\mu).
\ea
Here the new soft function $\S(\mu) =  J(\mu)\bar J(\mu)S(\mu)$ is the combined collinear and soft graphs and is well defined at NLO in pure dimensional regularization as seen in \cite{Cheung:2009sg} and \sec{previous}.  This new soft function is also infrared finite, as shown by using offshellness to regulate the infrared divergences of the collinear and soft graphs \cite{Cheung:2009sg}.  By running the functions between $\mu_H=Q$ and $\mu_\S=\sqrt{y_c}Q=\mu_{S,J}$ the \kt\ dijet rate \eqn{trueR} is reproduced to NLL accuracy. 

By choosing to run along the particular path in \fig{running}, it is clear that only the combined collinear and soft graphs are required for NLL summation.  Along this path, the general form of the $\nu$ anomalous dimension \eqn{gammanu} becomes
\ba
	\gamma_S^\nu(\mo)=\al_s(\mo)\sum_{m\geq0}\g_{S}^{(m)}\al_s^{\,m}(\mo),
\ea
which contains no large logarithms.  For N$^k$LL accuracy only the $m\leq k$ terms are required.  However, in general the $\g_S^{(0)}$ term, which is required for NLL accuracy, vanishes as seen in the \kt\ dijet rate above and all the cases in \cite{Chiu:2012ir}.  For N$^2$LL accuracy, therefore, only the hard running and the $\g_S^{(1)}$ term are required.

\bfig
   \centering
   \includegraphics[width=0.5\columnwidth]{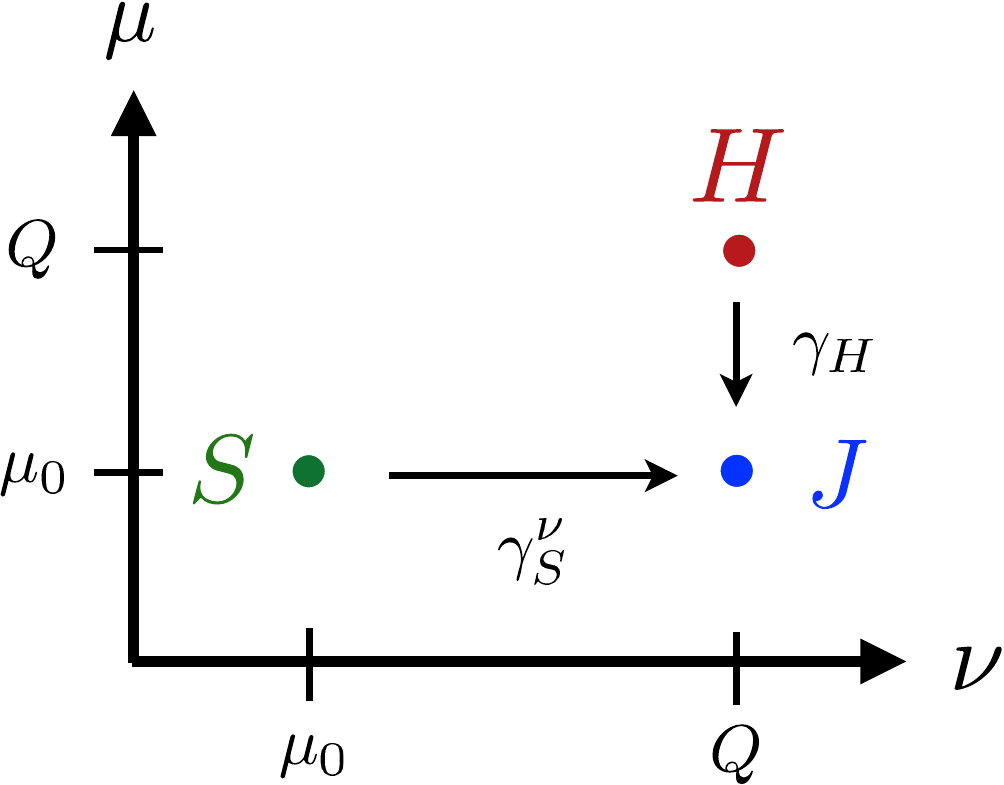} 
   \caption{Running the functions along a particular path.  Note that, up to NLL accuracy, the summed result is independent of the path chosen and the final $(\mu,\nu)$ point.}
   \label{fig:running}
\efig

\section{Conclusion}\label{sec:conc}

We have studied the (exclusive) \kt\ dijet rate using effective theory methods, and shown how to reproduce the coherent branching formalism result to NLL accuracy.  We must use the rapidity regulator if we wish to separate the NLO rate into regularized jet and soft functions. We have demonstrated how to sum to LL and NLL accuracy using the rapidity regulator in a path independent way, which can be generalized to any process that has a factorization theorem. We comment that the rapidity regulator is unnecessary for summing the large logarithms to NLL accuracy in the example of the \kt\ dijet rate. The same accuracy can be achieved if we consider the combined jet and soft function and run to the common jet and soft scale. We also find that using SCET with a rapidity regulator does not account for clustering effects and cannot improve the coherent branching formalism result.  A more complicated SCET-like theory may be able to properly account for these clustering effects, however, we do not explore such a theory in this paper.

\begin{acknowledgments}
We would like to thank B. Burrington, J-Y. M. Chiu, T. Dodds, C. Lee, M. Luke and S. Zuberi for helpful conversations.  This work was supported by the Natural Sciences and Engineering Research Council of Canada. WMYC was supported by the Ontario Graduate Scholarship.  SMF thanks the Institute for Nuclear Theory at the University of Washington for its hospitality and the U.S. Department of Energy for partial support during the completion of this work.
\end{acknowledgments}

\appendix*
\section*{Appendix}
\setcounter{equation}{0}

\begin{figure}[t]
   \centering
   \includegraphics[width=0.5\columnwidth]{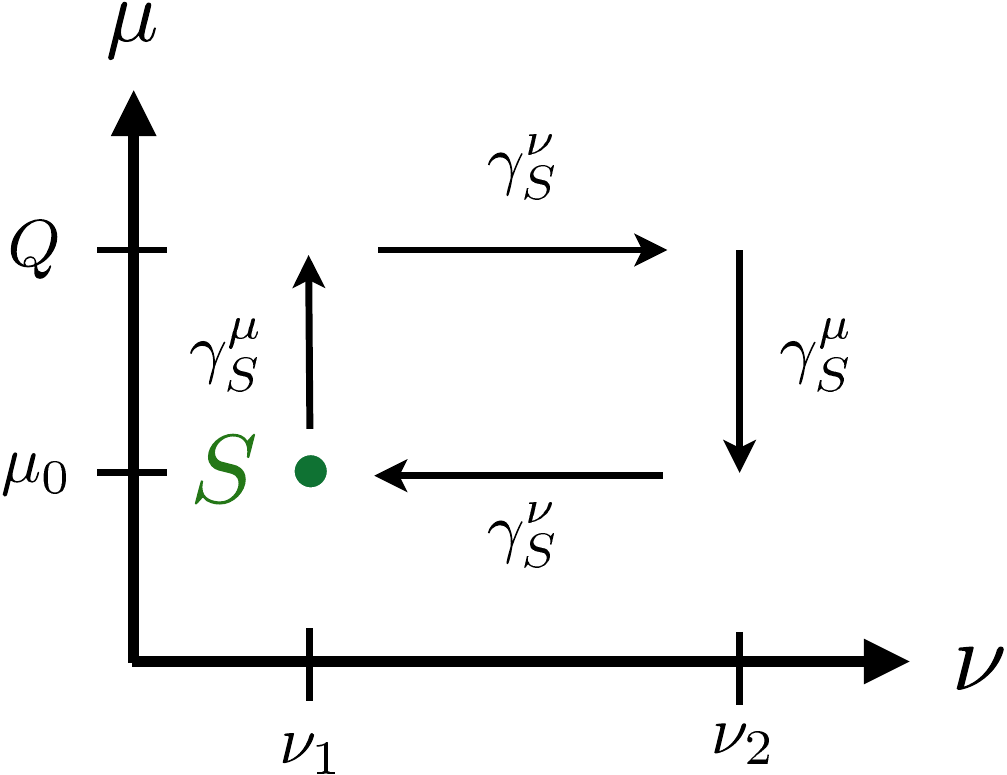} 
   \caption{Running the soft function in a rectangular path back to itself.  Naively this results in a LL phase when $\nu_2\gg\nu_1$.}
   \label{fig:softcircle}
\efig   

Here we show, using the soft function as an example, how to obtain \eqn{gammanu}, which allows us to sum to NLL accuracy. Our argument relies on factorization of scales, the consistency condition \eqn{gammacondition}, the vanishing curl \eqn{curl}, the general form of $\gamma_H$, and that the anomalous dimensions are defined perturbatively in $\al_s$. 

Factorization means that the anomalous dimensions of each function are sensitive only to scales relevant to it.  Therefore, the $\nu$ dependence of $\vec\gamma_J$ and $\vec\gamma_S$ will only be of the form $\log(\nu/Q)$ and $\log(\nu/\mo)$ respectively.  The consistency condition \eqn{gammacondition} requires that all $\nu$ dependence of $\vec\gamma_J$ and $\vec\gamma_S$ must cancel to all orders in perturbation theory.  As $\gamma_H^\mu$ is cusp-like and $\gamma_H^\nu$ vanishes, $\gamma_F^\mu$ can have at most a linear dependence on $\log(\nu/\nu_F)$ and $\gamma_F^\nu$ must have no $\nu$ dependence.  

The appearance of $\ln(\mu/\mu_0)$ in $\vec\gamma_F$, on the other hand, is not constrained. These logarithms can show up in arbitrary powers, as long as  they cancel one another in the sum $\vec\gamma_S + 2\vec\gamma_J$ to reproduce $\vec\gamma_H$. Fortunately these logarithms have negligible effect on NLL summation. This fact is made clear by the particular path shown in \fig{running}, where these logarithms vanish in $\gamma_S^\nu(\mo)$. Because of this and path independence, we suppress these terms in \eqn{generalgamma}.

The fact that $\gamma_S^\nu$ is independent of $\nu$ to all orders in $\al_s$ is also fixed by the form of $\gamma_S^\mu$ in \eqn{generalgamma} and the vanishing curl \eqn{curl}. Taking this general form and applying $\nu(d/d\nu)$ to both sides of \eqn{curl} yields
\ba\label{nonu}
	\mu\frac{d}{d\mu}\left(\nu\frac{d}{d\nu}\gamma_S^\nu(\mu,\nu)\right) = 0.
\ea
This means that $\nu(d/d\nu)\gamma^\nu_S(\mu,\nu)$ is independent of $\mu$ and in particular $\al_s(\mu)$. Such terms do not exist in perturbation theory, unless $\gamma_S^\nu(\mu,\nu)$ is independent of $\nu$.

The full $\mu$ dependence of $\gamma_S^\nu$ can therefore be obtained from $\gamma_S^\mu$ via integrating \eqn{curl}:
\ba\label{gnuapp}
	\gamma^\nu_S(\mu)=\gamma^\nu_S(\mu')+\int_{\mu'}^{\mu} \frac{d\mu''}{\mu''}\left(\nu\frac{d}{d\nu}\gamma_S^\mu(\mu'',\nu)\right).
\ea
In \eqn{gammanu} we choose $\mu'=\mo$ such that all logarithms in $\gamma^\nu_S(\mu')$ vanish and only the non-logarithmic terms remain. If \eqn{gnuapp} is not used, then running the soft function in the closed path shown in \fig{softcircle} would result in a large phase that spoils the LL accuracy of the results when $\nu_2\gg\nu_1$.

\bibliography{bibliography}

\begin{thebibliography}{24}%
\makeatletter
\providecommand \@ifxundefined [1]{%
 \@ifx{#1\undefined}
}%
\providecommand \@ifnum [1]{%
 \ifnum #1\expandafter \@firstoftwo
 \else \expandafter \@secondoftwo
 \fi
}%
\providecommand \@ifx [1]{%
 \ifx #1\expandafter \@firstoftwo
 \else \expandafter \@secondoftwo
 \fi
}%
\providecommand \natexlab [1]{#1}%
\providecommand \enquote  [1]{``#1''}%
\providecommand \bibnamefont  [1]{#1}%
\providecommand \bibfnamefont [1]{#1}%
\providecommand \citenamefont [1]{#1}%
\providecommand \href@noop [0]{\@secondoftwo}%
\providecommand \href [0]{\begingroup \@sanitize@url \@href}%
\providecommand \@href[1]{\@@startlink{#1}\@@href}%
\providecommand \@@href[1]{\endgroup#1\@@endlink}%
\providecommand \@sanitize@url [0]{\catcode `\\12\catcode `\$12\catcode
  `\&12\catcode `\#12\catcode `\^12\catcode `\_12\catcode `\%12\relax}%
\providecommand \@@startlink[1]{}%
\providecommand \@@endlink[0]{}%
\providecommand \url  [0]{\begingroup\@sanitize@url \@url }%
\providecommand \@url [1]{\endgroup\@href {#1}{\urlprefix }}%
\providecommand \urlprefix  [0]{URL }%
\providecommand \Eprint [0]{\href }%
\providecommand \doibase [0]{http://dx.doi.org/}%
\providecommand \selectlanguage [0]{\@gobble}%
\providecommand \bibinfo  [0]{\@secondoftwo}%
\providecommand \bibfield  [0]{\@secondoftwo}%
\providecommand \translation [1]{[#1]}%
\providecommand \BibitemOpen [0]{}%
\providecommand \bibitemStop [0]{}%
\providecommand \bibitemNoStop [0]{.\EOS\space}%
\providecommand \EOS [0]{\spacefactor3000\relax}%
\providecommand \BibitemShut  [1]{\csname bibitem#1\endcsname}%
\let\auto@bib@innerbib\@empty
\bibitem [{\citenamefont {Collins}\ \emph {et~al.}(1988)\citenamefont
  {Collins}, \citenamefont {Soper},\ and\ \citenamefont
  {Sterman}}]{Collins:1989gx}%
  \BibitemOpen
  \bibfield  {author} {\bibinfo {author} {\bibfnamefont {J.~C.}\ \bibnamefont
  {Collins}}, \bibinfo {author} {\bibfnamefont {D.~E.}\ \bibnamefont {Soper}},
  \ and\ \bibinfo {author} {\bibfnamefont {G.~F.}\ \bibnamefont {Sterman}},\
  }\href@noop {} {\bibfield  {journal} {\bibinfo  {journal}
  {Adv.Ser.Direct.High Energy Phys.}\ }\textbf {\bibinfo {volume} {5}},\
  \bibinfo {pages} {1} (\bibinfo {year} {1988})},\ \bibinfo {note} {publ. in
  `Perturbative QCD' (A.H. Mueller, ed.) (World Scientific Publ., 1989)},\
  \Eprint {http://arxiv.org/abs/hep-ph/0409313} {arXiv:hep-ph/0409313 [hep-ph]}
  \BibitemShut {NoStop}%
\bibitem [{\citenamefont {Catani}\ \emph {et~al.}(1993)\citenamefont {Catani},
  \citenamefont {Trentadue}, \citenamefont {Turnock},\ and\ \citenamefont
  {Webber}}]{Catani:1992ua}%
  \BibitemOpen
  \bibfield  {author} {\bibinfo {author} {\bibfnamefont {S.}~\bibnamefont
  {Catani}}, \bibinfo {author} {\bibfnamefont {L.}~\bibnamefont {Trentadue}},
  \bibinfo {author} {\bibfnamefont {G.}~\bibnamefont {Turnock}}, \ and\
  \bibinfo {author} {\bibfnamefont {B.}~\bibnamefont {Webber}},\ }\href
  {\doibase 10.1016/0550-3213(93)90271-P} {\bibfield  {journal} {\bibinfo
  {journal} {Nucl.Phys.}\ }\textbf {\bibinfo {volume} {B407}},\ \bibinfo
  {pages} {3} (\bibinfo {year} {1993})}\BibitemShut {NoStop}%
\bibitem [{\citenamefont {Catani}(1991)}]{Catani:1991gn}%
  \BibitemOpen
  \bibfield  {author} {\bibinfo {author} {\bibfnamefont {S.}~\bibnamefont
  {Catani}},\ }\href@noop {} {\bibfield  {journal} {\bibinfo  {journal} {Erice
  Proceedings, `QCD at 200-TeV', 21-41}\ } (\bibinfo {year}
  {1991})}\BibitemShut {NoStop}%
\bibitem [{\citenamefont {Catani}\ \emph {et~al.}(1991)\citenamefont {Catani},
  \citenamefont {Dokshitzer}, \citenamefont {Olsson}, \citenamefont {Turnock},\
  and\ \citenamefont {Webber}}]{Catani:1991hj}%
  \BibitemOpen
  \bibfield  {author} {\bibinfo {author} {\bibfnamefont {S.}~\bibnamefont
  {Catani}}, \bibinfo {author} {\bibfnamefont {Y.~L.}\ \bibnamefont
  {Dokshitzer}}, \bibinfo {author} {\bibfnamefont {M.}~\bibnamefont {Olsson}},
  \bibinfo {author} {\bibfnamefont {G.}~\bibnamefont {Turnock}}, \ and\
  \bibinfo {author} {\bibfnamefont {B.}~\bibnamefont {Webber}},\ }\href
  {\doibase 10.1016/0370-2693(91)90196-W} {\bibfield  {journal} {\bibinfo
  {journal} {Phys.Lett.}\ }\textbf {\bibinfo {volume} {B269}},\ \bibinfo
  {pages} {432} (\bibinfo {year} {1991})}\BibitemShut {NoStop}%
\bibitem [{\citenamefont {Brown}\ and\ \citenamefont
  {Stirling}(1990)}]{Brown:1990nm}%
  \BibitemOpen
  \bibfield  {author} {\bibinfo {author} {\bibfnamefont {N.}~\bibnamefont
  {Brown}}\ and\ \bibinfo {author} {\bibfnamefont {W.}~\bibnamefont
  {Stirling}},\ }\href {\doibase 10.1016/0370-2693(90)90502-W} {\bibfield
  {journal} {\bibinfo  {journal} {Phys.Lett.}\ }\textbf {\bibinfo {volume}
  {B252}},\ \bibinfo {pages} {657} (\bibinfo {year} {1990})}\BibitemShut
  {NoStop}%
\bibitem [{\citenamefont {Brown}\ and\ \citenamefont
  {Stirling}(1992)}]{Brown:1991hx}%
  \BibitemOpen
  \bibfield  {author} {\bibinfo {author} {\bibfnamefont {N.}~\bibnamefont
  {Brown}}\ and\ \bibinfo {author} {\bibfnamefont {W.}~\bibnamefont
  {Stirling}},\ }\href {\doibase 10.1007/BF01559740} {\bibfield  {journal}
  {\bibinfo  {journal} {Z.Phys.}\ }\textbf {\bibinfo {volume} {C53}},\ \bibinfo
  {pages} {629} (\bibinfo {year} {1992})}\BibitemShut {NoStop}%
\bibitem [{\citenamefont {Dissertori}\ and\ \citenamefont
  {Schmelling}(1995)}]{Dissertori:1995qx}%
  \BibitemOpen
  \bibfield  {author} {\bibinfo {author} {\bibfnamefont {G.}~\bibnamefont
  {Dissertori}}\ and\ \bibinfo {author} {\bibfnamefont {M.}~\bibnamefont
  {Schmelling}},\ }\href {\doibase 10.1016/0370-2693(95)01174-O} {\bibfield
  {journal} {\bibinfo  {journal} {Phys.Lett.}\ }\textbf {\bibinfo {volume}
  {B361}},\ \bibinfo {pages} {167} (\bibinfo {year} {1995})}\BibitemShut
  {NoStop}%
\bibitem [{\citenamefont {Banfi}\ \emph {et~al.}(2002)\citenamefont {Banfi},
  \citenamefont {Salam},\ and\ \citenamefont {Zanderighi}}]{Banfi:2001bz}%
  \BibitemOpen
  \bibfield  {author} {\bibinfo {author} {\bibfnamefont {A.}~\bibnamefont
  {Banfi}}, \bibinfo {author} {\bibfnamefont {G.}~\bibnamefont {Salam}}, \ and\
  \bibinfo {author} {\bibfnamefont {G.}~\bibnamefont {Zanderighi}},\
  }\href@noop {} {\bibfield  {journal} {\bibinfo  {journal} {JHEP}\ }\textbf
  {\bibinfo {volume} {0201}},\ \bibinfo {pages} {018} (\bibinfo {year}
  {2002})},\ \Eprint {http://arxiv.org/abs/hep-ph/0112156}
  {arXiv:hep-ph/0112156 [hep-ph]} \BibitemShut {NoStop}%
\bibitem [{\citenamefont {Cheung}\ \emph {et~al.}(2009)\citenamefont {Cheung},
  \citenamefont {Luke},\ and\ \citenamefont {Zuberi}}]{Cheung:2009sg}%
  \BibitemOpen
  \bibfield  {author} {\bibinfo {author} {\bibfnamefont {W.~M.-Y.}\
  \bibnamefont {Cheung}}, \bibinfo {author} {\bibfnamefont {M.}~\bibnamefont
  {Luke}}, \ and\ \bibinfo {author} {\bibfnamefont {S.}~\bibnamefont
  {Zuberi}},\ }\href {\doibase 10.1103/PhysRevD.80.114021} {\bibfield
  {journal} {\bibinfo  {journal} {Phys. Rev.}\ }\textbf {\bibinfo {volume}
  {D80}},\ \bibinfo {pages} {114021} (\bibinfo {year} {2009})},\ \Eprint
  {http://arxiv.org/abs/0910.2479} {arXiv:0910.2479 [hep-ph]} \BibitemShut
  {NoStop}%
\bibitem [{\citenamefont {Bauer}\ \emph {et~al.}(2000)\citenamefont {Bauer},
  \citenamefont {Fleming},\ and\ \citenamefont {Luke}}]{Bauer:2000ew}%
  \BibitemOpen
  \bibfield  {author} {\bibinfo {author} {\bibfnamefont {C.~W.}\ \bibnamefont
  {Bauer}}, \bibinfo {author} {\bibfnamefont {S.}~\bibnamefont {Fleming}}, \
  and\ \bibinfo {author} {\bibfnamefont {M.~E.}\ \bibnamefont {Luke}},\ }\href
  {\doibase 10.1103/PhysRevD.63.014006} {\bibfield  {journal} {\bibinfo
  {journal} {Phys. Rev.}\ }\textbf {\bibinfo {volume} {D63}},\ \bibinfo {pages}
  {014006} (\bibinfo {year} {2000})},\ \Eprint
  {http://arxiv.org/abs/hep-ph/0005275} {arXiv:hep-ph/0005275 [hep-ph]}
  \BibitemShut {NoStop}%
\bibitem [{\citenamefont {Bauer}\ \emph {et~al.}(2001)\citenamefont {Bauer},
  \citenamefont {Fleming}, \citenamefont {Pirjol},\ and\ \citenamefont
  {Stewart}}]{Bauer:2000yr}%
  \BibitemOpen
  \bibfield  {author} {\bibinfo {author} {\bibfnamefont {C.~W.}\ \bibnamefont
  {Bauer}}, \bibinfo {author} {\bibfnamefont {S.}~\bibnamefont {Fleming}},
  \bibinfo {author} {\bibfnamefont {D.}~\bibnamefont {Pirjol}}, \ and\ \bibinfo
  {author} {\bibfnamefont {I.~W.}\ \bibnamefont {Stewart}},\ }\href {\doibase
  10.1103/PhysRevD.63.114020} {\bibfield  {journal} {\bibinfo  {journal} {Phys.
  Rev.}\ }\textbf {\bibinfo {volume} {D63}},\ \bibinfo {pages} {114020}
  (\bibinfo {year} {2001})},\ \Eprint {http://arxiv.org/abs/hep-ph/0011336}
  {arXiv:hep-ph/0011336 [hep-ph]} \BibitemShut {NoStop}%
\bibitem [{\citenamefont {Bauer}\ and\ \citenamefont
  {Stewart}(2001)}]{Bauer:2001ct}%
  \BibitemOpen
  \bibfield  {author} {\bibinfo {author} {\bibfnamefont {C.~W.}\ \bibnamefont
  {Bauer}}\ and\ \bibinfo {author} {\bibfnamefont {I.~W.}\ \bibnamefont
  {Stewart}},\ }\href {\doibase 10.1016/S0370-2693(01)00902-9} {\bibfield
  {journal} {\bibinfo  {journal} {Phys. Lett.}\ }\textbf {\bibinfo {volume}
  {B516}},\ \bibinfo {pages} {134} (\bibinfo {year} {2001})},\ \Eprint
  {http://arxiv.org/abs/hep-ph/0107001} {arXiv:hep-ph/0107001 [hep-ph]}
  \BibitemShut {NoStop}%
\bibitem [{\citenamefont {Bauer}\ \emph
  {et~al.}(2002{\natexlab{a}})\citenamefont {Bauer}, \citenamefont {Pirjol},\
  and\ \citenamefont {Stewart}}]{Bauer:2001yt}%
  \BibitemOpen
  \bibfield  {author} {\bibinfo {author} {\bibfnamefont {C.~W.}\ \bibnamefont
  {Bauer}}, \bibinfo {author} {\bibfnamefont {D.}~\bibnamefont {Pirjol}}, \
  and\ \bibinfo {author} {\bibfnamefont {I.~W.}\ \bibnamefont {Stewart}},\
  }\href {\doibase 10.1103/PhysRevD.65.054022} {\bibfield  {journal} {\bibinfo
  {journal} {Phys. Rev.}\ }\textbf {\bibinfo {volume} {D65}},\ \bibinfo {pages}
  {054022} (\bibinfo {year} {2002}{\natexlab{a}})},\ \Eprint
  {http://arxiv.org/abs/hep-ph/0109045} {arXiv:hep-ph/0109045 [hep-ph]}
  \BibitemShut {NoStop}%
\bibitem [{\citenamefont {Bauer}\ \emph
  {et~al.}(2002{\natexlab{b}})\citenamefont {Bauer}, \citenamefont {Fleming},
  \citenamefont {Pirjol}, \citenamefont {Rothstein},\ and\ \citenamefont
  {Stewart}}]{Bauer:2002nz}%
  \BibitemOpen
  \bibfield  {author} {\bibinfo {author} {\bibfnamefont {C.~W.}\ \bibnamefont
  {Bauer}}, \bibinfo {author} {\bibfnamefont {S.}~\bibnamefont {Fleming}},
  \bibinfo {author} {\bibfnamefont {D.}~\bibnamefont {Pirjol}}, \bibinfo
  {author} {\bibfnamefont {I.~Z.}\ \bibnamefont {Rothstein}}, \ and\ \bibinfo
  {author} {\bibfnamefont {I.~W.}\ \bibnamefont {Stewart}},\ }\href {\doibase
  10.1103/PhysRevD.66.014017} {\bibfield  {journal} {\bibinfo  {journal} {Phys.
  Rev.}\ }\textbf {\bibinfo {volume} {D66}},\ \bibinfo {pages} {014017}
  (\bibinfo {year} {2002}{\natexlab{b}})},\ \Eprint
  {http://arxiv.org/abs/hep-ph/0202088} {arXiv:hep-ph/0202088 [hep-ph]}
  \BibitemShut {NoStop}%
\bibitem [{\citenamefont {Freedman}\ and\ \citenamefont
  {Luke}(2012)}]{Freedman:2011kj}%
  \BibitemOpen
  \bibfield  {author} {\bibinfo {author} {\bibfnamefont {S.~M.}\ \bibnamefont
  {Freedman}}\ and\ \bibinfo {author} {\bibfnamefont {M.}~\bibnamefont
  {Luke}},\ }\href {\doibase 10.1103/PhysRevD.85.014003} {\bibfield  {journal}
  {\bibinfo  {journal} {Phys.Rev.}\ }\textbf {\bibinfo {volume} {D85}},\
  \bibinfo {pages} {014003} (\bibinfo {year} {2012})},\ \Eprint
  {http://arxiv.org/abs/1107.5823} {arXiv:1107.5823 [hep-ph]} \BibitemShut
  {NoStop}%
\bibitem [{\citenamefont {Hornig}\ \emph {et~al.}(2009)\citenamefont {Hornig},
  \citenamefont {Lee},\ and\ \citenamefont {Ovanesyan}}]{Hornig:2009vb}%
  \BibitemOpen
  \bibfield  {author} {\bibinfo {author} {\bibfnamefont {A.}~\bibnamefont
  {Hornig}}, \bibinfo {author} {\bibfnamefont {C.}~\bibnamefont {Lee}}, \ and\
  \bibinfo {author} {\bibfnamefont {G.}~\bibnamefont {Ovanesyan}},\ }\href
  {\doibase 10.1088/1126-6708/2009/05/122} {\bibfield  {journal} {\bibinfo
  {journal} {JHEP}\ }\textbf {\bibinfo {volume} {05}},\ \bibinfo {pages} {122}
  (\bibinfo {year} {2009})},\ \Eprint {http://arxiv.org/abs/0901.3780}
  {arXiv:0901.3780 [hep-ph]} \BibitemShut {NoStop}%
\bibitem [{\citenamefont {Ellis}\ \emph {et~al.}(2010)\citenamefont {Ellis},
  \citenamefont {Vermilion}, \citenamefont {Walsh}, \citenamefont {Hornig},\
  and\ \citenamefont {Lee}}]{Ellis:2010rwa}%
  \BibitemOpen
  \bibfield  {author} {\bibinfo {author} {\bibfnamefont {S.~D.}\ \bibnamefont
  {Ellis}}, \bibinfo {author} {\bibfnamefont {C.~K.}\ \bibnamefont
  {Vermilion}}, \bibinfo {author} {\bibfnamefont {J.~R.}\ \bibnamefont
  {Walsh}}, \bibinfo {author} {\bibfnamefont {A.}~\bibnamefont {Hornig}}, \
  and\ \bibinfo {author} {\bibfnamefont {C.}~\bibnamefont {Lee}},\ }\href
  {\doibase 10.1007/JHEP11(2010)101} {\bibfield  {journal} {\bibinfo  {journal}
  {JHEP}\ }\textbf {\bibinfo {volume} {1011}},\ \bibinfo {pages} {101}
  (\bibinfo {year} {2010})},\ \Eprint {http://arxiv.org/abs/1001.0014}
  {arXiv:1001.0014 [hep-ph]} \BibitemShut {NoStop}%
\bibitem [{\citenamefont {Chiu}\ \emph {et~al.}(2011)\citenamefont {Chiu},
  \citenamefont {Jain}, \citenamefont {Neill},\ and\ \citenamefont
  {Rothstein}}]{Chiu:2011qc}%
  \BibitemOpen
  \bibfield  {author} {\bibinfo {author} {\bibfnamefont {J.-y.}\ \bibnamefont
  {Chiu}}, \bibinfo {author} {\bibfnamefont {A.}~\bibnamefont {Jain}}, \bibinfo
  {author} {\bibfnamefont {D.}~\bibnamefont {Neill}}, \ and\ \bibinfo {author}
  {\bibfnamefont {I.~Z.}\ \bibnamefont {Rothstein}},\ }\href@noop {} {\
  (\bibinfo {year} {2011})},\ \Eprint {http://arxiv.org/abs/1104.0881}
  {arXiv:1104.0881 [hep-ph]} \BibitemShut {NoStop}%
\bibitem [{\citenamefont {Chiu}\ \emph {et~al.}(2012)\citenamefont {Chiu},
  \citenamefont {Jain}, \citenamefont {Neill},\ and\ \citenamefont
  {Rothstein}}]{Chiu:2012ir}%
  \BibitemOpen
  \bibfield  {author} {\bibinfo {author} {\bibfnamefont {J.-y.}\ \bibnamefont
  {Chiu}}, \bibinfo {author} {\bibfnamefont {A.}~\bibnamefont {Jain}}, \bibinfo
  {author} {\bibfnamefont {D.}~\bibnamefont {Neill}}, \ and\ \bibinfo {author}
  {\bibfnamefont {I.~Z.}\ \bibnamefont {Rothstein}},\ }\href@noop {} {\
  (\bibinfo {year} {2012})},\ \Eprint {http://arxiv.org/abs/1202.0814}
  {arXiv:1202.0814 [hep-ph]} \BibitemShut {NoStop}%
\bibitem [{\citenamefont {Kelley}\ \emph {et~al.}(2012)\citenamefont {Kelley},
  \citenamefont {Walsh},\ and\ \citenamefont {Zuberi}}]{Kelley:2012kj}%
  \BibitemOpen
  \bibfield  {author} {\bibinfo {author} {\bibfnamefont {R.}~\bibnamefont
  {Kelley}}, \bibinfo {author} {\bibfnamefont {J.~R.}\ \bibnamefont {Walsh}}, \
  and\ \bibinfo {author} {\bibfnamefont {S.}~\bibnamefont {Zuberi}},\
  }\href@noop {} {\  (\bibinfo {year} {2012})},\ \Eprint
  {http://arxiv.org/abs/1202.2361} {arXiv:1202.2361 [hep-ph]} \BibitemShut
  {NoStop}%
\bibitem [{\citenamefont {Bauer}\ and\ \citenamefont
  {Schwartz}(2007)}]{Bauer:2006mk}%
  \BibitemOpen
  \bibfield  {author} {\bibinfo {author} {\bibfnamefont {C.~W.}\ \bibnamefont
  {Bauer}}\ and\ \bibinfo {author} {\bibfnamefont {M.~D.}\ \bibnamefont
  {Schwartz}},\ }\href {\doibase 10.1103/PhysRevD.76.074004} {\bibfield
  {journal} {\bibinfo  {journal} {Phys.Rev.}\ }\textbf {\bibinfo {volume}
  {D76}},\ \bibinfo {pages} {074004} (\bibinfo {year} {2007})},\ \Eprint
  {http://arxiv.org/abs/hep-ph/0607296} {arXiv:hep-ph/0607296 [hep-ph]}
  \BibitemShut {NoStop}%
\bibitem [{\citenamefont {Manohar}(2003)}]{Manohar:2003vb}%
  \BibitemOpen
  \bibfield  {author} {\bibinfo {author} {\bibfnamefont {A.~V.}\ \bibnamefont
  {Manohar}},\ }\href {\doibase 10.1103/PhysRevD.68.114019} {\bibfield
  {journal} {\bibinfo  {journal} {Phys.Rev.}\ }\textbf {\bibinfo {volume}
  {D68}},\ \bibinfo {pages} {114019} (\bibinfo {year} {2003})},\ \Eprint
  {http://arxiv.org/abs/hep-ph/0309176} {arXiv:hep-ph/0309176 [hep-ph]}
  \BibitemShut {NoStop}%
\bibitem [{\citenamefont {Manohar}\ and\ \citenamefont
  {Stewart}(2007)}]{Manohar:2006nz}%
  \BibitemOpen
  \bibfield  {author} {\bibinfo {author} {\bibfnamefont {A.~V.}\ \bibnamefont
  {Manohar}}\ and\ \bibinfo {author} {\bibfnamefont {I.~W.}\ \bibnamefont
  {Stewart}},\ }\href {\doibase 10.1103/PhysRevD.76.074002} {\bibfield
  {journal} {\bibinfo  {journal} {Phys. Rev.}\ }\textbf {\bibinfo {volume}
  {D76}},\ \bibinfo {pages} {074002} (\bibinfo {year} {2007})},\ \Eprint
  {http://arxiv.org/abs/hep-ph/0605001} {arXiv:hep-ph/0605001 [hep-ph]}
  \BibitemShut {NoStop}%
\bibitem [{\citenamefont {Georgi}(1993)}]{Georgi:1994qn}%
  \BibitemOpen
  \bibfield  {author} {\bibinfo {author} {\bibfnamefont {H.}~\bibnamefont
  {Georgi}},\ }\href@noop {} {\bibfield  {journal} {\bibinfo  {journal}
  {Ann.Rev.Nucl.Part.Sci.}\ }\textbf {\bibinfo {volume} {43}},\ \bibinfo
  {pages} {209} (\bibinfo {year} {1993})}\BibitemShut {NoStop}%
\end{thebibliography}%

\end{document}